\documentclass[runningheads]{llncs}
\usepackage{graphicx}
\usepackage{hyperref}
\usepackage{booktabs}
\usepackage{tabularx}
\usepackage{tikz}
\def\checkmark{\tikz\fill[scale=0.4](0,.35) -- (.25,0) -- (1,.7) -- (.25,.15) -- cycle;} 
\usepackage{graphicx}
\usepackage{amsmath}
\usepackage{floatrow}
\usepackage{subfig}
\usepackage{comment}
\usepackage{multirow}
\usepackage{url}

\usepackage{svg}
\usepackage{pdfpages}

\usepackage[most]{tcolorbox}

\newtcolorbox{mytextbox}[1][]{%
  sharp corners,
  enhanced,
  colback=white,
  attach title to upper,
  #1
}

\newcommand{\ToolName}[0]{\texttt{Domainator}}
\newcommand{\TotalCount}[0]{7}
\newcommand{\MalwareCount}[0]{5}
\newcommand{\ToolsCount}[0]{2}
\newcommand{\cdpNA}[0]{\textcolor{gray}{n/a}}

\begin{document}
\title{\ToolName{}: Detecting and Identifying DNS-Tunneling Malware Using Metadata Sequences}

\titlerunning{Domainator: Detecting and Identifying DNS-Tunneling Malware}

\author{Denis~Petrov\inst{1}\orcidID{0009-0000-7131-6844}
\and
Pascal~Ruffing\inst{1}\orcidID{0009-0001-3540-983X}
\and
Sebastian~Zillien\inst{2}\orcidID{0000-0003-3360-1251}
\and
Steffen~Wendzel\inst{2}\orcidID{0000-0002-1913-5912}
}

\authorrunning{D.~Petrov \emph{et al.}}

\institute{Worms University of Applied Sciences, Germany\\
\email{\{petrov,ruffing\}@hs-worms.de}\\
\and
Ulm University, Germany\\
\email{\{firstname.lastname\}@uni-ulm.de}
}
\maketitle              
\begin{abstract}
In recent years, malware with tunneling (or: covert channel) capabilities is on the rise. While malware research led to several methods and innovations, the detection and \emph{differentiation} of malware solely based on its DNS tunneling features is still in its infancy. Moreover, no work so far has used the DNS tunneling traffic to gain knowledge over the current \emph{actions} taken by the malware.

In this paper, we present \ToolName{}, an approach to detect and differentiate state-of-the-art malware and DNS tunneling tools without relying on trivial (but quickly altered) features such as ``magic bytes'' that are embedded into subdomains. Instead, we apply an analysis of sequential patterns to identify specific types of malware. We evaluate our approach with \TotalCount{} different malware samples and tunneling tools and can identify the particular malware based on its DNS traffic. We further infer the rough \emph{behavior} of the particular malware through its DNS tunneling artifacts. Finally, we compare our \ToolName{} with related methods.


\end{abstract}

\begin{mytextbox}[colupper=blue,fontupper=\bfseries\normalsize]
This is a pre-print. The final version of this paper will be published in the proceedings of the \emph{20th International Conference on Availability, Reliability and Security} (ARES'25).
\end{mytextbox}

\section{Introduction}
Tunneling data through a protocol that is otherwise not meant for such activities provides malware authors with the capabilities of transporting stolen data and subsequent command and control messages to and from a C2 server in a hidden manner. Their focus on evading traditional detection methods allows the malware and its communication channel(s) to remain undiscovered. As such tunnels break the security policy of their environments in a stealthy manner, literature refers to these channels as \emph{covert channels} \cite{zander2007survey}. More than 100 recent cases of malware employing covert channels have been summarized by Strachanski \emph{et al.} \cite{Strachanski:StegomalwareSurvey} (2024), Knöchel and Karius \cite{KnoechelKarius:TextStegomalware24} (2024) as well as Caviglione and Mazurczyk \cite{NeverMindMalware} (2022).

Network tunneling exploits features of common Internet protocols \cite{NeverMindMalware}, with DNS being the most frequently used protocol for this purpose \cite{Strachanski:StegomalwareSurvey}. While the detection of network-specific covert channels was studied for years, there is still a lack regarding the \emph{differentiation} -- or: \emph{identification} -- of malware. Moreover, studies on detecting or identifying network covert channels usually target \emph{purely academic} implementations.

This paper aims to address this gap by showing that we can not only \emph{detect} but \emph{identify} (i.e., separate) \emph{real-world} DNS-based malware, and also understand its \emph{behavior}. In particular, our contributions are the following:

\begin{enumerate}
    \item We introduce a methodology for the identification of \emph{real-world} malware samples on the network-level;
    
    \item Targeting DNS-based tunneling, we are able to differentiate multiple malware samples solely based on statistical features of their subdomain utilization, that we feed into a Random Forest classifier;

    \item We show that our classification approach can be used to detect malware \emph{behavior}, i.e., we can tell which action a malware performs;
    
    \item We further show that we can identify malware samples even when their communication pattern is slightly modified;
    
    \item We provide our dataset to the scientific community.
\end{enumerate}

The remainder of the paper is structured as follows. We cover fundamentals and related work in Sect.~\ref{sect:relwrk} and present our methodology in Sect.~\ref{sect:meth}. We evaluate our approach in Sect.~\ref{sect:eval}, which also introduces the analyzed malware samples. Sect.~\ref{sect:disc} discusses our methodology while Sect.~\ref{sect:concl} concludes.

\section{Related Work and Fundamentals}\label{sect:relwrk}

\subsection{Related Work}
\paragraph{Establishment of DNS Tunnels}
A plethora of methods exist to \emph{embed} covert messages into network traffic, see, e.g., the following survey publications \cite{csur,zander2007survey,mileva2014covert}. Recent findings unveiled that the predominant protocol utilized by malware is DNS \cite{Strachanski:StegomalwareSurvey}. Summarized, to hide secret information in DNS traffic, attackers exploit the header fields of the DNS protocol and the values of the DNS entries that are attached to the header (these are called \emph{resource records}). 

\paragraph{Detection of DNS Tunnels}
Several papers study the \emph{detection} of network covert channels, including DNS-based ones, see, e.g., \cite{csur,NIHbook} 
for comprehensive overviews.
The related work on detecting DNS-based covert channels and tunneling barely addresses real-world malware traffic and almost exclusively focuses on academic tools and methods. Moreover, it targets the detection (and not the identification) of these tools. 
Known approaches for detecting DNS-based covert channels and tunneling are the following: Gao \emph{et al.} \cite{GraphTunnel} aims to detect DNS-based tunnels using a framework called GraphTunnel, which leverages graph neural networks to model DNS recursive resolution processes as graphs. By aggregating node and edge features with the GraphSage framework, GraphTunnel achieves 100\% accuracy in detecting both known and unknown DNS tunnels, even in challenging environments with wildcard DNS where it maintains an F1-score of 99.78\%. In addition to detecting tunnels, GraphTunnel accurately identifies DNS tunneling tools with a success rate of more than 98.57\%. In contrast, Born and Gustafson focus on character frequency analysis to detect DNS tunnels~\cite{born2010detectingdnstunnelsusing}. Their approach uses unigram, bigram, and trigram models to detect anomalies in the frequency patterns, based on the observation that domain names typically follow Zipf’s law. This method effectively identifies DNS tunnels by highlighting deviations from expected character frequency distributions. With a focus on machine learning, Buczak \emph{et al.}\ \cite{buczak} develop a method for detecting DNS tunnels by applying Random Forest classifiers to analyze features extracted from PCAP data. The authors curated a dataset by extracting relevant features from DNS traffic, such as domain name length, query types, and packet length, enabling them to effectively distinguish between normal DNS activity and tunneling activity. Their approach achieved a detection accuracy of over 95\%, 
demonstrating the effectiveness of Random Forest models in identifying DNS tunnels. Similarly, in the work of Alkasassbeh and Almseidin~\cite{alkasassbeh}, the detection of DNS tunneling using machine learning techniques is explored, with a particular focus on evaluating the effectiveness of different classifiers. In their experiments, Random Forest classifiers again produced the best results, achieving a precision of 98\% in detecting DNS tunneling traffic. Žiža \emph{et al.} have also shown an approach based on Random Forests that has an accuracy of 99.7\%. Another recent study has outlined detection approaches for both network-based malware communication and other forms of media, summarizing them in a comprehensive overview \cite{Survey-detection}. While these works target the bare detection of DNS tunnels, our work additionally aims to identify the particular malware that generated the observed DNS traffic. Moreover, we attempt to determine the behavior of the malware solely by analyzing its network communication, in regards to which we found no related works.

\paragraph{Anomaly Detection Using Sequences}
Chandola \emph{et al.} provide a comprehensive survey on anomaly detection in discrete sequences \cite{Chandola}. The authors categorize existing research into three main approaches: sequence-based anomaly detection, which focuses on identifying entire sequences that deviate from the norm; subsequence-based anomaly detection, which involves detecting anomalous subsequences within longer sequences; and pattern frequency-based anomaly detection, which identifies anomalies based on unusual frequency patterns of specific sequences. The survey provides a detailed review of techniques for each of these formulations, offering insights into how different approaches can be applied across various domains. In \cite{Loganathan}, Loganathan \emph{et al.} introduce a sequence-to-sequence (Seq2Seq) pattern learning algorithm for real-time anomaly detection in network traffic. The authors propose a multi-attribute model that predicts sequences of TCP packets based on prior sequences using a Seq2Seq encoder-decoder architecture. The model effectively learns the expected order of packets, allowing it to identify deviations that signal potential anomalies or intrusions, achieving an accuracy of 97\% for detecting anomalous TCP packets.

\subsection{Fundamentals on DNS-utilization by Malware}
Due to the crucial nature of DNS for a successful communication on the Internet, it is also an attractive target for threat groups and their malware, with the aim of keeping their communication inconspicuous.

\begin{figure}
    \centering    
    \includegraphics[width=0.95\columnwidth]{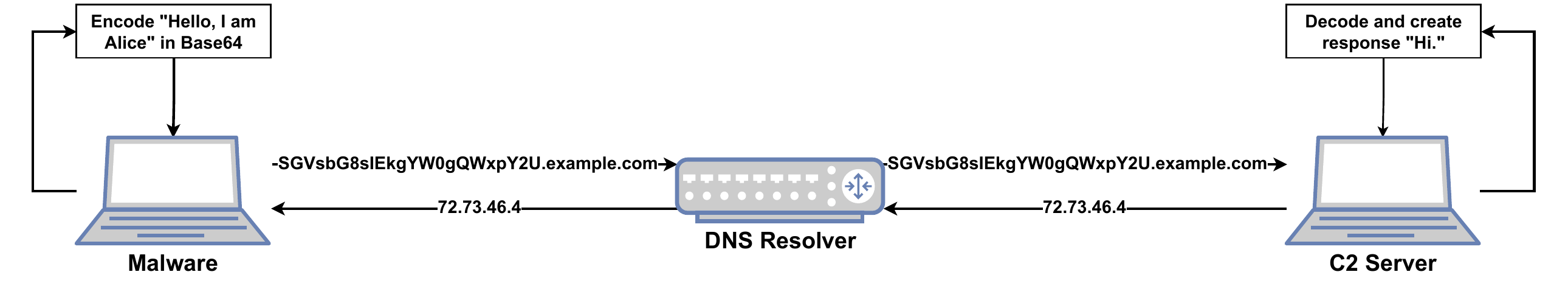}
    \caption{A visualization of the malware instance communicating with its C2 server}
    \label{fig:c2-communication}
\end{figure}

As DNS does not support arbitrary payload transfer natively, malicious actors utilize its request and response structure by having a malware issue DNS queries for a domain registered and controlled by them. The data transferred by the client-side of the communication channel is inserted as a subdomain string, which can then be read and interpreted by a remote C2 server. Such communication is indirect, with a DNS resolver standing between the client and the server, with the communication loop being visualized in Fig.~\ref{fig:c2-communication}. When the length of the data, which the malware attempts to exfiltrate, exceeds either a pre-defined value or the maximum allowed size for a DNS packet, it is split into multiple smaller chunks that are sent individually, usually coupled with an offset value to indicate which part of the message is transferred. In order to keep the domains valid and the requests less suspicious, the plain-text data often has undergone some form of encoding or encryption. Depending on the malware implementation, the record types chosen for the DNS queries may vary. On the C2 server, the request is processed, a fitting response is created and again encoded into a meaningful string for the resource record type used. This may be an IP address in the case of record types \textit{A} and \textit{AAAA}, or another subdomain for \textit{CNAME}.

\section{Framework and Dataset}\label{sect:meth}

\subsection{The \ToolName{} DNS Covert Channel Detection Framework}
To gather the data necessary for our evaluation, we have constructed an isolated physical testbed environment on which we can observe the analyzed malware samples. The testbed consists of a Windows- and Linux-based computer where the malware is executed, a Linux-based computer that acts as a DNS resolver, as well as the host for the respective C2 server in a separate module, and a router connecting the computers into a local network.

For the execution of each malware sample, the client-side device was reset to the initial state it had before any malware infection took place. In addition to that, no debugging or malware analysis tools were present on the system at the same time as the malware to prevent any external influences. Since our goal is to analyze the original network traffic generated by the malware, we made sure that it was not recognized or influenced by any local antivirus and malware detection tools. To this end, we deactivated the in-built antivirus software (e.g., Windows Defender), since some of the used samples were easily detected. This enables an unobstructed communication that we can build our DNS-based malware identification upon. 

On the server-side, an instance of the open-source tool DNSChef \cite{DNSChef} was utilized as a DNS resolver. Based on the domain name in the DNS query, DNSChef has been configured to either respond with a dummy IP address or to forward the packet to an address the C2 server can listen on. When a packet is captured by the server, it interprets the request and returns an appropriate answer, which is also handled by the resolver. The list of domains to be forwarded is compiled from domains we have observed during the operation of the malware samples in Sect.~\ref{subsect:samples}. This allows the simulation of an authentic network communication for both malicious and non-malicious requests.

A limiting factor for the reproduction of the malicious communication is the lack of knowledge regarding the hardware or software, as well as their exact implementation, that the threat group used originally with the associated sample. In order to replicate this command and control setup, we have either recreated the C2 server by examining the malware and its traffic, and determining the expected responses, or have used already available C2 server implementations. One such implementation we leveraged is Saitama, which was publicly available for our analysis~\cite{C2-Saitama}. Every C2 server listens for traffic coming from the DNS resolver with a destination port \textit{53} and parses for the domains utilized by the respective malware instance. The network packets are then processed by filtering the subdomains, based on the record type and the current malware requirements. Due to the distinctive nature of each malware, the procedures of the server in this step are highly individual. Following, a fitting reply is constructed and sent to the client as a DNS response.

\subsection{Selection and Description of DNS Malware and Tunneling Tools}\label{subsect:samples}
To cover various DNS data exfiltration methods and their use in malicious scenarios, we have chosen a list of seven tools: five DNS-utilizing malwares, including two different implementations of the same malware, as well as two open-source tools, at least one of which has been utilized in malicious campaigns:

\textit{RogueRobin} is a malware created by the Iranian-linked DarkHydrus threat group \cite{DarkHydrus}, targeting government figures within the Middle East. There are two known versions of the malware, the main differences between being the delivered payload: either a PowerShell-based script or a C\# executable \cite{RogueRobin}. They also differ in the approach used for encoding the data that is sent through the DNS communication channel. We have analyzed both versions of the malware and created a C2 server for each so that we can explore their capabilities.\\
Initially, the \textit{PowerShell version} runs various checks to test whether it is being run within a virtual environment or a debugger \cite{RogueRobin}. If those tests are passed, it begins rotating between the DNS record types stored in a hard-coded list and sends a DNS query for each one, in an attempt to find out which records can be used successfully for communication. The very first request is also utilized by the malware as a way to register itself with the C2 server. It inserts the process ID as a \textit{base64} encoded string into the subdomain of the query, and expects to get an integer value back, that is used as a unique ID \cite{RogueRobin}. If the response to any of the requests was not successful or valid, the malware would repeat the query, before moving to the next record. After each resource type has been tested, an evaluation string is sent back to the server \cite{RogueRobin-Ironnet}, indicating which records were successful. The complete transfer of this string is considered by the malware as a job and is designated with an ID, that is sent alongside the encoded data. Longer data strings are split into chunks that are sent separately, in which case the job ID aids to rebuild the original full message. A sample DNS request, which transfers data, has the following structure: \textit{uniqueID}\textbar{}-\textbar{}\textit{jobID}\textbar{}-\textbar{}\textit{dataOffset}\textbar{}\textit{isMoreFlag\textbar{}}-\textbar{}\textit{encodedData}.\textit{example}.\textit{com}. The malware also gathers and transfers information about the victim system and the user, before querying the C2 server for any available jobs.\\
The \textit{.NET version} executes a similar initial sequence utilizing a round-robin resource record rotation, with the list of record types having a different order \cite{RogueRobin-Ironnet}. The data in this version uses a hexadecimal encoding, which is then fully converted into letters through a simple, hard-coded substitution alphabet. The malware also includes a character at the beginning of each covert message, indicating the current mode, i.e., data transfer. When sending data back to the server, .NET RogueRobin separates the hidden data into two parts, one containing meta information about the data and the other being the data itself. To distinguish between each part, a single character from a pre-defined list is inserted. An example packet that sends data from the victim would have the form of \textit{mode}\textbar{}\textit{uniqueID}\textbar{}\textit{jobID}\textbar{}\textit{dataOffset}\textbar{}\textit{isMoreFlag}\textbar{}\textit{sepChar}\textbar{}\textit{encodedData}.\textit{example}.\textit{com}. For non-data transferring queries, the covert message would also include a single hard-coded character either at the end or behind the \textit{jobID} position.

\textit{Saitama} is a malware that was targeted towards Jordanian government officials. It was first analyzed by Malwarebytes \cite{Saitama}, which is the main source for our initial analysis. On the command and control side, we have employed a publicly available C2 server implementation \cite{C2-Saitama}. In comparison to \textit{RogueRobin}, \textit{Saitama} establishes longer pauses between each sent packet and utilizes a form of base36 encoding with a mixed character positions to introduce randomness in the resulting string. An additional preventative measure to reduce the similarity of each malware run is the internal counter that chooses a random number upon execution, and is then utilized in the data encoding. The initial request \textit{Saitama} makes is a registration query that requests an ID it can be identified by later. The ID itself is sent back as the last octet of an IPv4 address, while other responses use each octet to encode the covert response. Subsequently, the malware retrieves a job from the C2 server, which can either be from a list of pre-defined commands or a custom one that will take more DNS queries to be acquired. Once the job is complete, the sample sends the result back to the server in chunks. In the case of no available jobs, it initiates a form of a keep-alive connection which sends a burst of packets in an interval of six to eight hours.

\textit{Symbiote} is a type of malware that specifically targets Linux-based systems and was first analyzed by Intezer, which our description is mainly based on~\cite{Symbiote}. Its main target has been the infiltration of banks in Latin America. Unlike other typical malware, \textit{Symbiote} is not a standalone executable, instead, being a shared object file that inserts itself into running processes. To achieve this, it uses the \textit{LD\_PRELOAD} environment variable to ensure that it is loaded first in every new process \cite{Mitre-DLH}.
There are five known versions of \textit{Symbiote}, with only two implementing their own DNS covert communication protocol. A later version of the malware uses a modified version of the DNS tunneling tool DNSCat2, which we will discuss later. The primary objective of \textit{Symbiote} is providing remote access to the compromised system and stealing credentials by capturing them from SSH or SCP processes. It uses two different communication approaches. For exfiltration, the data is encrypted, encoded into hexadecimal format and embedded into DNS \textit{A}-record queries using the following format: 
\textit{packetNumber\textbar{}.\textbar{}machineID\textbar{}.\textbar{}hexEncPayload.example.com}.
The packet number begins at $11111$ and is incremented for each sent chunk. The machine ID, made up of data from the \emph{uname} syscall, helps to differentiate between infected systems. \\
The second communication method is used to download scripts from its C2 server for further attacks. To accomplish this, it uses a similar approach, with the utilized resource record being changed to \textit{TXT}. It uses the following query format:
\textit{packetNumber\textbar{}.\textbar{}machineID.example.com}.
For this mode, the packet number starts at 0, which indicates that the malware requests the ED25519 signature of the script. This is later used to verify the downloaded data and prevent the malware from executing scripts from other sources.
The packet number is then incremented, signaling the actual start of the download. As the size of DNS packets is limited, the script must be split into chunks and transmitted individually. The server can determine the respective chunk through the packet number in the requests. The selected chunk is encrypted and sent in the DNS response.

\textit{DNSCat2} is an open-source remote control tool, which utilizes DNS tunneling~\cite{DNSCat2}. It strongly focuses on command and control functionality like file upload or download and tunneling shell sessions. DNSCat2 offers two operation modes, a ``typical'' connection through a local DNS Server and a direct connection.
The direct connection uses UDP on port 53 and mimics DNS traffic to some extent but will not hold up against a proper inspection.
The ``typical'' mode uses actual DNS requests that are sent through a local DNS server.
Similar to other tools, DNSCat2 offers various configuration options to tailor the behavior to different networking environments. By default, it uses encrypted connections between client and server, and a single server can interface with multiple client simultaneously.
The complexity of DNSCat2 is also observed in the communication protocol itself, which possesses TCP-like components, e.g., sequence numbers and acknowledgments. Internal commands of DNSCat2 follow another protocol that works on top of the low-level protocol. The binary data is transported through the DNS requests/responses as a hex-encoded string.

\textit{Symbiote DNSCat2}~\cite{Symbiote} is a modified DNSCat2 version of Symbiote, presumably customized by the same attack group. The changes mainly concern how the tool can be started. The reverse-engineered binary revealed that the attackers removed the argument parameters of the main function and created a custom \textit{argv} variable with the following fixed parameters: \texttt{--no-encryption --dns domain=git.bancodobrasil.dev,type=TXT}. These are used to initialize the tool and set up a connection to the C2 server through \textit{TXT} record messages. The transferred data is not being encrypted, instead applying a base64 encoding.

\textit{Iodine}~\cite{Iodine} is a general-purpose tunneling tool that can transport IPv4 traffic through the DNS protocol, offering a vast and versatile set of parameters and settings. The main use case is to gain Internet access in a network where general Internet access is blocked, but DNS traffic is still allowed through the firewall. This could be used to circumvent captive portals in public Wi-Fi networks or to exfiltrate information from corporate networks.
As it is a tunneling software, it focuses on bidirectional communication, session stability and throughput. The general concept of the tunneling is similar to the samples mentioned above.

\subsection{Generation of Data for the Evaluation}\label{subsect:data-gen}
In a real attack scenario, the DNS communication would be hidden within a multitude of non-malicious traffic. When evaluating our malware identification framework, we must take that into consideration. However, due to the testbed having no Internet access, it is not possible to have ordinary, user-generated traffic directly mixed with the malware traffic. We resolve this by expanding the dataset with legitimate traffic, as described in Sect.~\ref{subsect:dataset-expansion}.

We have chosen a total of seven ``scenarios'' where the malware samples are executed and the traffic produced by them is recorded. The scenarios are meant to cover different aspects of the communication and to gain an ample overview of the malware behavior. Every possible combination of a malware and scenario has been created (as long as the feature covered by the scenario is part of the capabilities of the malware). To have a consistent and comparable dataset across all of our malware samples, some of the recordings were conducted with a set time limit that is based on the recorded scenario. The scenarios are:
\begin{enumerate}
    \item Handshake (HS) - observes the start-up sequences initiated by the malware. In this mode the custom-made C2 server is available and ready to respond to any requests, thus the malware can register and display its behavior in a realistic infection scenario. Due to each malware having a unique initial chain of requests, we have designated all traffic produced up to the first Idle/Keep Alive request, as the Handshake of the malware. For that reason, the recordings have contrasting lengths and sizes, as well as no fixed duration. 
    \item Handshake with Fake Internet (HS Fakenet) - inspects the traffic generated by the malware when the C2 server is unavailable or unreachable. The mode intends to mimic a realistic infection scenario, and therefore a fake Internet connection was established between the victim and our uplink server. While the malware will be unable to register with a legitimate C2 server, we are interested in the behavior and frequency of the requests sent by the infected device. The recordings are limited to \textit{30~minutes}.
    \item Handshake Offline (HS Offline) - a scenario similar to the Handshake with Fake Internet, where no network connection is simulated, therefore none of the requests sent by the victim device are answered. While there is no change in the C2 server accessibility from the previous mode, this scenario allows us to observe whether the behavior of the malware undergoes any changes when a system is completely offline. Equivalent to the Handshake with Fake Internet, the recordings were done over a \textit{30~minutes} time period.
    \item Idle/Keep Alive - a mode that enables the C2 server and observes the communication after the initial Handshake. This means a connection with the malware has been established but no jobs will be sent to the sample once the handshake is complete. Due to some malware samples implementing pause intervals to make their traffic less recognizable, we have decided each recording of this mode to have a \textit{12~hours} length.
    \item Steal SSH key through upload (UL SSH) - a plausible infection scenario for a job which utilizes the malware to upload the contents of a file from the victim machine to the C2 server. To have a more representative set of data, we have decided to record this scenario with both a 4096 bit RSA and an ED25519 private keys. For consistency, we have used the exact same keys for each malware sample. There is no set time limit for the recordings of this scenario, instead storing all of the sent packets from the job acquisition to the last packet of the job.
    \item Steal Large File (UL File) - a scenario that utilizes the same malware capabilities as the Steal SSH one. The goal is to see how each sample handles a larger file, i.e., a Microsoft Office Word document. To do that, we first encode the document into base64, and then proceed with the upload. We used the same file for each sample, and the recording length varies between them.
    \item Download of a File (DL File) - the last scenario transfers a file from the C2 server to the infected device. Contrasting the previous upload modes, the malware sends requests which may contain metadata, rather than stolen information, so that the commanding server can interpret it and respond with segments of the file. In real infection circumstances, this file could be additional instructions, another stage of the malware, or a new malware altogether. We have chosen to use a network scanner tool, as it fits the use-case while being relatively light in size. Nevertheless, due to the limited space and frequencies of the requests, sending the full file may take extremely long time, so we limited the recordings to a maximum of \textit{12 hours}.
\end{enumerate}

Tab.~\ref{table:scenarios} shows all the combinations of malware and scenarios that we recorded. As the SSH scenario uploads two separate files, each has been stored in its own recording, making a total of 8 possible recordings for each tool or malware.

\begin{table*}
\centering
\caption{Malware and open-source tools, and the scenario combinations for our traffic recordings. \emph{*Dnscat2 has been used in a malware campaign}}
\resizebox{1.0\textwidth}{!}{
\begin{tabular}{r|c|c|c|c|c|c|c}\toprule
    \textbf{Malware and Tools} & \textbf{Handshake} & \textbf{HS Fakenet} & \textbf{HS Offline} & \textbf{Keep Alive} & \textbf{UL SSH} & \textbf{UL File} & \textbf{DL File} \\ \midrule
    RogueRobin PS & \checkmark & \checkmark & \checkmark & \checkmark & \checkmark & \checkmark & \checkmark \\
    RogueRobin .NET & \checkmark & \checkmark & \checkmark & \checkmark & \checkmark & \checkmark & \checkmark \\
    Saitama & \checkmark & \checkmark & \checkmark & \checkmark & \checkmark & \checkmark & \checkmark \\
    Symbiote & - & - & - & - & \checkmark & \checkmark & \checkmark \\
    Symbiote DNSCat2 & \checkmark & \checkmark & \checkmark & \checkmark & \checkmark & \checkmark & \checkmark \\ \midrule
    DNSCat2* & \checkmark & \checkmark & \checkmark & \checkmark & \checkmark & \checkmark & \checkmark \\
    Iodine & \checkmark & \checkmark & \checkmark & \checkmark & \checkmark & \checkmark & \checkmark \\ \bottomrule
\end{tabular}
}
\label{table:scenarios}
\end{table*}

\subsection{Expanding the Collected Dataset}\label{subsect:dataset-expansion}
While the recorded network traffic covers a broad range of the potential tunneling usage, taking each recording as a single sample is not sufficient for a meaningful analysis. This is due to the low total number of samples available to training and testing our methodology, as we would be grouping all requests from the same domain. In addition, it is not an efficient approach that can be applied to a live-environment as the full recording may take a long period of time to be captured. For this reason, we have taken an approach that combines all of the recordings into a single set of data and then processes $n$ packets under the same domain in a sliding window. The packets in this combined dataset follow a consistent scenario sequence and have the same order as they have been sent in. Likewise, the packet order in each window follows the original order of the traffic recording. This also leads to some windows containing traffic from two scenarios, which would mirror an authentic classification attempt in the wild. 

We have analyzed various window size configurations and did not notice a major difference in the results. For this reason, we have chosen a size of $10$ packets per window, as this allows \ToolName{} to not only be used \emph{post factum}, but also in a live setting with a manageable number of requests that are necessary to do the identification. As there may not always be a full window of $10$ requests, especially in the case of legitimate domains, we have set a minimum of $3$ requests per window that are necessary for the calculation of the statistical features.

Although the order of the scenarios does not influence our results, we have built a plausible attack configuration which takes the following arrangement:
\begin{center}
    Handshake Offline $\rightarrow$ Handshake Fakenet $\rightarrow$ Handshake $\rightarrow$ \\ Download $\rightarrow$ Idle $\rightarrow$ Steal RSA $\rightarrow$ Steal ED25519 $\rightarrow$ Steal  File
\end{center}
 
The recordings done with the live malware contain all the additional legitimate packets that were sent by other services during the execution. However, the malware produces a much higher amount of traffic in comparison to the rest of the applications, thus building an imbalanced ratio of malicious and non-malicious traffic. 

To have more equal proportions between these, we have added legitimate, non-malicious DNS requests from a dataset recorded in the span of a day on a DNS resolver managed by an Internet service provider \cite{Ziza2023}. The authors have anonymized the timestamps and the true user IP addresses, but have kept the adjustments consistent to the real traffic, i.e., all requests from a single user are mapped to the same new address. Additionally, two exfiltration tools were executed during the recorded period -- \emph{Iodine} and \emph{DNSExfiltrator}. The dataset also recognizes the presence of benign data exfiltration performed by two distinct antivirus tools, namely \emph{Eset} and \emph{McAfee}. All exfiltration requests have been marked as such, so that they can be separated from the rest of the data, and were used by us as part of the validation in Sect.~\ref{subsect:validation}.

In addition to that, we have created a validation set consisting of recordings of our malware samples that were not utilized in the training. These recordings follow a similar malware-scenario pairing, but for each one of them a different parameter was changed or the code itself was modified to simulate possible adjustments an adversary could undertake to evade a detection algorithm. These changes include new files being uploaded or downloaded, the removal of used domain names, changes in the used resource records, changes in the encoding algorithm and alphabet, and reduction or increase of the communication capacity per DNS request. The goal of this set is to validate the identification process and its robustness to distortions in the traffic.

\subsection{Feature Selection}
When analyzing any of the given malware samples, a fairly simple approach would be to search for re-occurring fixed strings which uniquely identify the sample (e.g., subdomains which always begin with a certain substring). We call such values \textit{magic bytes} and consider their utilization an insufficient solution because they could be changed by the malware authors by adjusting a single variable. Doing detection over specific anomalies can also lead to the incorrect classification of anomalous-looking traffic originating from legitimate sources, e.g., antivirus software. Moreover, it is not possible to reliably identify such magic bytes in most cases.

\begin{figure}
    \centering
    \includegraphics[width=1\columnwidth]{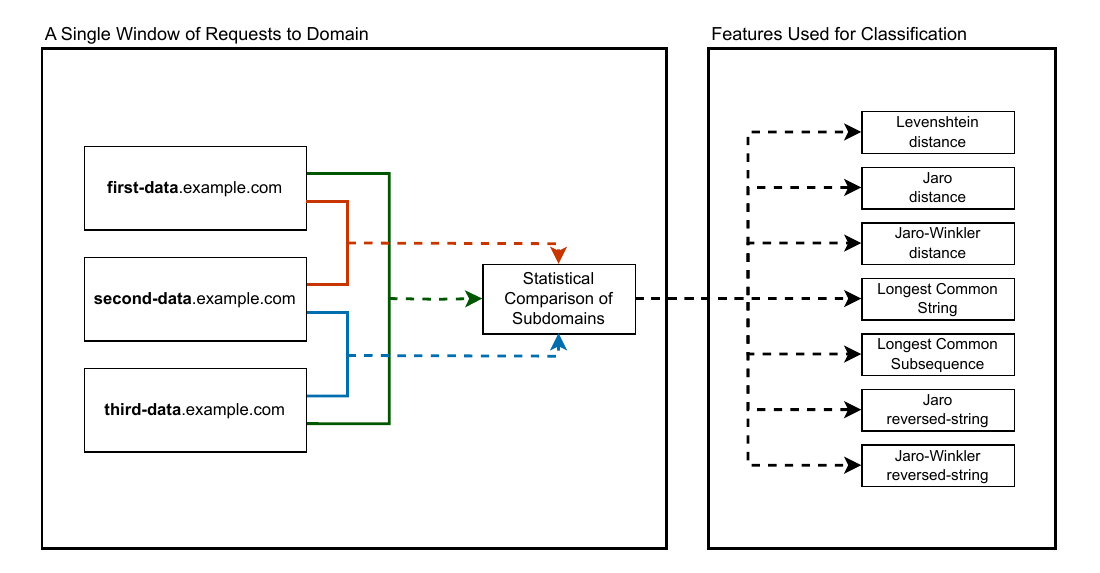}
    \caption{A single window of three requests and their combination into the statistical metrics. Each metric consists of the mean value for the window.}
    \label{fig:methodology-structure}
\end{figure}

We decided to work independent of such magic bytes and have concentrated our efforts on creating a set of statistical values which capture the overall characteristics of the traffic generated by the malware, instead of the individual requests. When detecting tunneling presence in network flows, the metrics used by the papers described in Sect.~\ref{sect:relwrk} have already shown to be effective, as they search for anomalies not usually present in legitimate traffic. However, when attempting to do identification, there is an overlap of the anomalous characteristics, as each sample is malicious. Using our windowing approach, we analyze a snippet of the traffic and build all pair combinations within an individual window. Following, various string metrics are calculated for every combination, resulting in a subdomain similarity score for each metric. We take the scores on a per metric basis, and determine the mean value, thus assigning similarity scores to the window itself, which act as the input features to our detection and identification classifiers. The string metric features are listed on the right side of Fig.~\ref{fig:methodology-structure}. Our approach has also been visualized in Fig.~\ref{fig:methodology-structure}, with the possible combinations of DNS requests being shown on the left side of the graphic, and the calculated mean scores resulting in the features on the right side.

The Levenshtein distance, as well as the Jaro and Jaro-Winkler similarities, provide data on the number of edits it takes to get from one string to the other. As each malware has a different implementation of its communication protocol, there is a contrast between the order of occurring changes and their execution in consecutive requests sent to the C2 server. It is also not unusual for the communication protocol to include a form of a victim ID, a job ID or a counter, which add some repetitiveness to the requests. The Jaro-Winkler similarity considers such reoccurring elements at the beginning of the strings by having a scaling component that results in a higher similarity score. To cover a broader comparison, we also calculate the Jaro and Jaro-Winkler scores for the reversed strings, so that subdomains ending in the same suffix can also be weighted correspondingly. Furthermore, it is possible that static patterns within the strings are separated by a chunk of dynamic data. The longest common subsequence covers such cases by matching two strings and finding equivalent segments, even with gaps of variable length. In contrast, the longest common substring searches for an uninterrupted pattern of matching characters, and in the case of multiple separated sequences, it considers only the longest one. The combination of these features results in traffic windows exhibiting similar subdomain patterns to be grouped together. These clusters are distinguishable from one another, which allows a classifier to correctly identify a future window without relaying on specific spatial and temporal variables, e.g., the timestamp or the timings between the packets.

Using the combination of the scenario recordings and the legitimate traffic, we built a dataset containing $43,212$ windows, with the total number of windows for each malware being presented in Tab.~\ref{table:counts}. The data was split into a train part and a test part, with a proportion of $80\%$ to $20\%$, which were subsequently passed to a simple classifier. Although we have analyzed various classification and regression methods, our approach showed the greatest potential with a Random Forest classification, due to its overall score and efficiency. In our final approach, we have trained multiple such classifiers, with all of them utilizing the same input features. The evaluation is split into several stages, which can either work separately or be put together into a pipeline. For all results, we have listed the scores of the Random Forest classifiers. 

\begin{table*}[h]
\centering
\caption{Number of windows in the full dataset, together with their distribution per action}
\begin{tabular}{r|r||r|r|r|r}\toprule
    \textbf{Type of Traffic} & \textbf{Total} & \textbf{Handshake} & \textbf{Keep Alive} & \textbf{Upload} & \textbf{Download} \\ \midrule
    Legitimate & 15,914 & - & - & - & - \\
    \midrule
    RogueRobin PS & 3,005 & 24 & 1,106 & 114 & 1,761 \\
    RogueRobin .NET & 3,043 & 23 & 930 & 524 & 1,566 \\
    Saitama & 242 & 8 & 6 & 158 & 70 \\
    Symbiote & 697 & - & - & 23 & 674 \\
    Symbiote DNSCat2 & 7,331 & 7 & 4,319 & 21 & 2,984 \\ \midrule
    DNSCat2 & 8,054 & 8 & 4,269 & 23 & 3,754 \\
    Iodine & 4,926 & 9 & 4,297 & 14 & 606 \\ \bottomrule
\end{tabular}
\label{table:counts}
\end{table*}

\section{Results and Evaluation}\label{sect:eval}

\subsection{Malware Detection} \label{sub:malware-detection}
\begin{figure}
\CenterFloatBoxes
\begin{floatrow}
\ttabbox
  {
    \renewcommand{\arraystretch}{1.4}
    \centering
    \begin{tabular}{r c c|r}
        & \multicolumn{2}{c|}{Predicted}\\ \midrule
        Actual & {Malic.} & {Legitim.} & \textbf{F1-Score} \\ \midrule
        Malicious & 5,201 & 259 & \textbf{0.955}\\
        Legitimate & 38 & 3,145 & \textbf{0.972}\\
        \midrule
        & & & \textbf{0.966} \\
    \end{tabular}
  }
  {\caption{Classification results for the malware detection}\label{tab:results-malware-detection}}
\killfloatstyle
\ffigbox
  {\includegraphics[width=1\columnwidth]{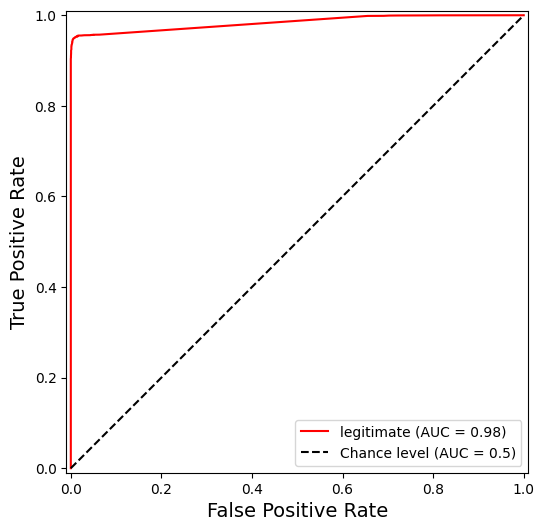}}
  {\caption{ROC curve: malware detection}\label{fig:auc-malware-detection}}
\end{floatrow}
\end{figure}

Our initial step was to conduct a binary classification, where all non-legitimate traffic was labeled as malicious. The goal of this was to examine whether we can create a detection algorithm that can differentiate between the malicious and non-malicious windows using the described features. The classifier had an overall accuracy (F1-Score) of 0.966, with a macro average precision of 0.958 and recall of 0.970. The false-positive rate lies at $1.2\%$. The distribution of the predicted labels is shown in Tab.~\ref{tab:results-malware-detection}, together with the associated ROC curve in Fig.~\ref{fig:auc-malware-detection}.

\subsection{Malware Identification} \label{sub:malware-ident}
\begin{figure}[htp]
    \centering
    \subfloat[Confusion matrix\label{fig:classification_matrix}]{%
      \includegraphics[width=0.483\columnwidth]{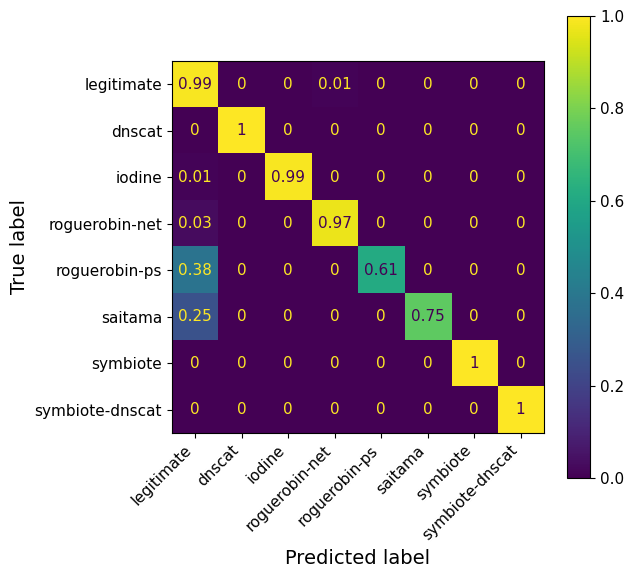}%
    }\hfil
    \subfloat[ROC curve\label{fig:classification_aoc}]{%
      \includegraphics[width=0.417\columnwidth]{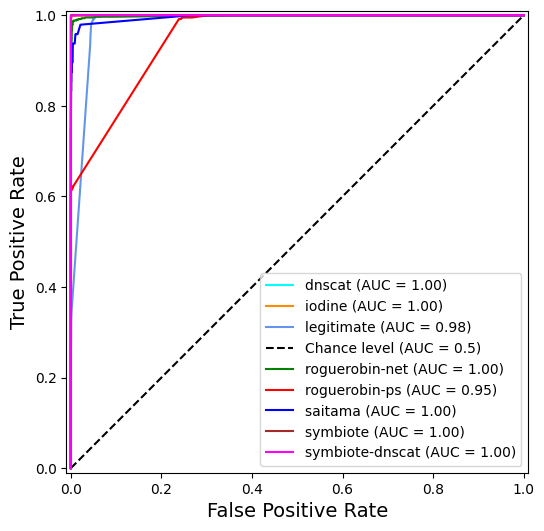}%
    }
    \caption{Malware identification results}
    \label{fig:malware_classification_subfig}
\end{figure}

The next step was to increase the granularity of the classification by having each malicious window be labeled as the malware that it originates from. This creates the broader task of not only detecting the presence of malicious DNS requests, but also separating and identifying each sample. The classifier achieved an F1-score of 0.964, similar to the accuracy of the previous binary approach, with the more precise scores being shown in Fig.~\ref{fig:classification_matrix}. The macro precision of the classifier is 0.982, while its recall is 0.911. Fig.~\ref{fig:classification_aoc} presents the ROC curves.

In both of these cases, there appears to be an outlier scenario that the classifier is not able to correctly detect and identify. The \textit{Idle} traffic produced by PowerShell RogueRobin continuously sends the exact same packet requesting a new job ID to be provided by the server. This leads to a statistical uniformity that is consistent with the behavior observed by a big portion of the non-malicious traffic. For the detection of the 259 windows falsely identified as legitimate in Tab.~\ref{tab:results-malware-detection}, 221 (85\%) can be attributed to this scenario. Although not impactful to our classifiers, due to the smaller number of windows, the \textit{Idle} windows of Saitama have the same communication pattern and are incorrectly identified as legitimate traffic.

\subsection{Scenario Identification} \label{sub:scenario-ident}
Subsequently, we implement identification of the behavior and actions the tools have undertaken. An action is defined as the function currently executed by the tool and correlates to the scenarios in Sect.~\ref{subsect:data-gen}. The traffic within our dataset can be categorized into four such actions: performing the initial handshake, being idle, downloading a file, and uploading a file. Although some of our designed scenarios cover the same action (e.g., multiple \textit{Uploads}), these can be joined into a singular group and therefore pose no obstruction to the classification. Non-malicious traffic performs none of these actions. The distribution of windows for each action has been described in Tab.~\ref{table:counts}.

An imbalance between the number of DNS requests for the actions can be observed, as it is ordinary that the malware sends more packets during the more data transfer intensive actions. While it can skew the score of the classifier, it is more authentic to a real infection scenario. However, a limitation resulting directly from this design choice is the very small number of DNS requests within the handshake action, due to the tools requiring to do it only once at the beginning of their interaction with the C2 server. As a result of this, we have decided to exclude this action from the scenario identification.

There are two approaches that can be used for the training of the classifier. The first one attempts to identify the window directly, without any previous steps being taken, while the second one assumes that the window has already been identified as malicious and removes the legitimate traffic from the dataset. While both approaches led to a similar accuracy of 0.857 and 0.876, the inclusion of the legitimate traffic in the first approach skews the results as the precision and recall of the respective classifier are 0.832 and 0.804, in comparison to 0.892 and 0.881 for the second one. The main difference between the two is caused by PowerShell RogueRobin and Saitama traffic being identified as non-malicious, similarly to the malware identification. 

\begin{figure}
    \centering
    \includegraphics[width=0.95\columnwidth]{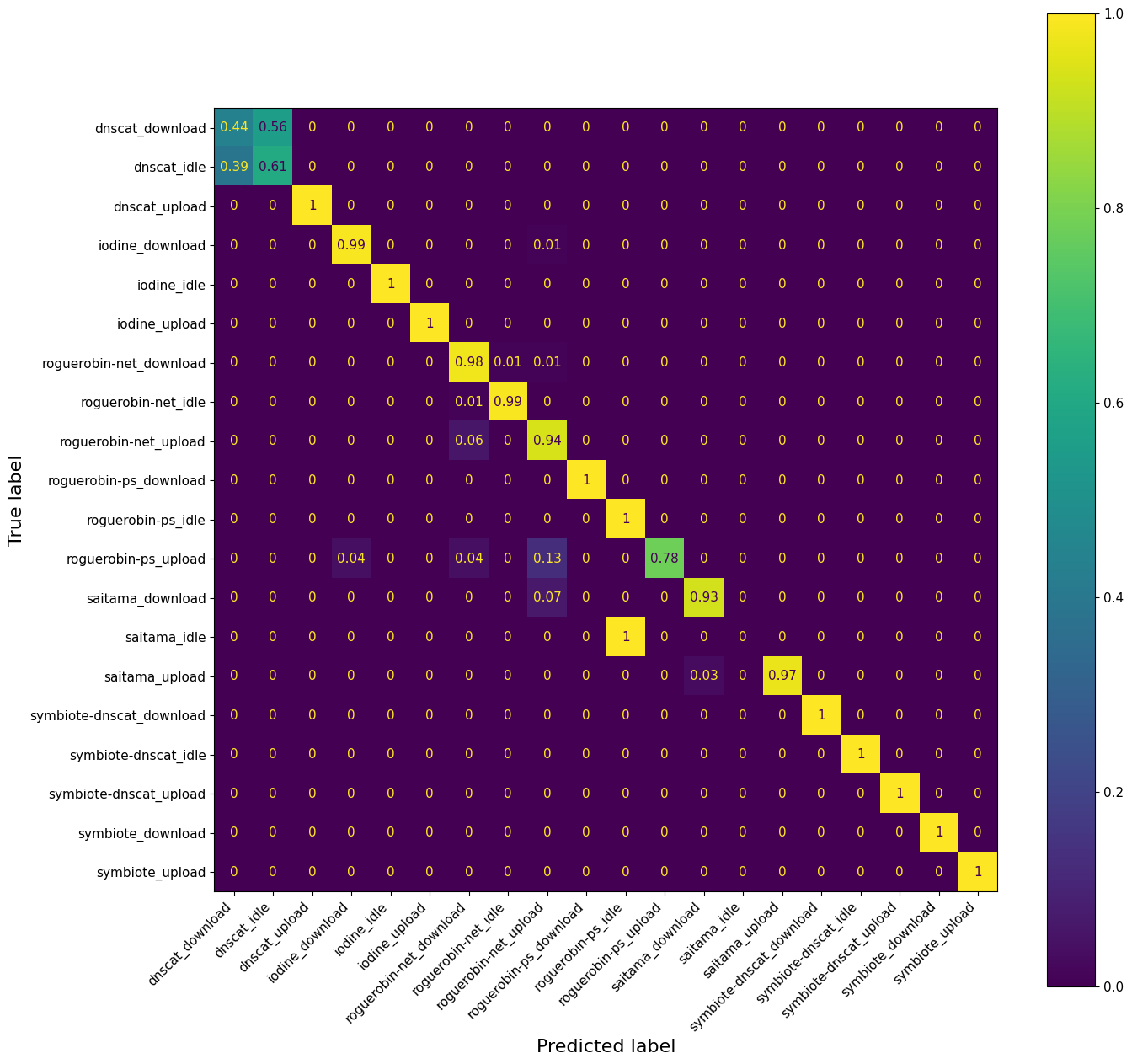}
    \caption{Confusion matrix with the scores of the scenario identification classifier}
    \label{fig:scenario-identification}
\end{figure}

The results of the identification done by the approach without legitimate traffic, as visualized in Fig.~\ref{fig:scenario-identification}, show that the classifier is struggling with two of the actions recorded with DNSCat2. The cause for this is the usage of the same patterns during the download of a file and idle operation which our statistical features cannot differentiate. Although the approach with no legitimate traffic correctly classifies the idle traffic of RogueRobin, a different malware sample (e.g., Saitama) producing the same unchanged DNS requests would lead to identical inaccuracies. A further refinement of the method could have the identification task be performed in two separate stages -- the first identifies the specific malware, and the second classifies the action.
\begin{figure}[htp]
    \centering
    \subfloat[Confusion matrix\label{fig:pure_scenario_matrix}]{%
      \includegraphics[width=0.483\columnwidth]{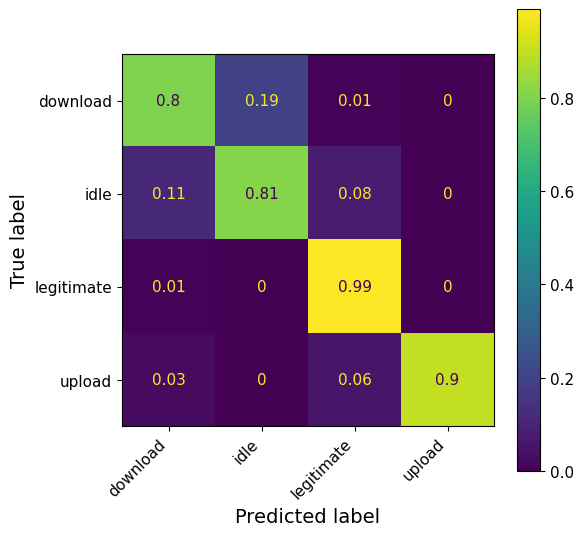}%
    }\hfil
    \subfloat[ROC curve\label{fig:pure_scenario_aoc}]{%
      \includegraphics[width=0.417\columnwidth]{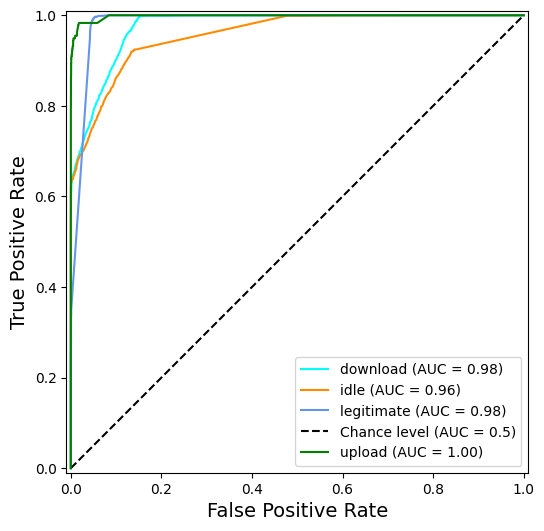}%
    }
    \caption{Generalized scenario identification results}
    \label{fig:pure-scenario-identification}
\end{figure}

We have also evaluated a more generalized scenario classification in which the malware is not taken into account but the classifier has to identify what action is being performed. As the training dataset mixes windows from different tools for the same actions, which means the used patterns are not complimentary to one another, it is performing worse than the more precise approaches. However, the goal of this form of classification would be to attempt identifying malicious actions taken by malware which is not in the dataset, but has a similar communication protocol. The results of this method are visualized in Fig.~\ref{fig:pure-scenario-identification}, and the overall accuracy is 0.878, with macro precision of 0.889 and recall of 0.878.

\subsection{Validation} \label{subsect:validation}
In addition to the classification of the initial dataset, we wanted to analyze how well the classifiers can manage traffic produced by the same malware and tools, but with adapted settings. These adapted settings are: different files being uploaded, altered domains and resource records, a modified set of used encoding characters, and a modified size of the data that is sent with each request. Although we have not crossed each tool with each setting variation, we have built an ample validation dataset which we passed to our trained classifiers.

The evaluation of this dataset was done by taking each PCAP file and computing the statistical windows, then letting the classifier make its prediction. Subsequently, an average score is calculated for the whole file. Tab.~\ref{table:validation-scores} shows most of the entries of this dataset, together with the averaged F1-scores for the malware and scenario classifiers. As the validation dataset contains overlaps of malware-scenario pairings with similar changes (e.g., Iodine Upload with multiple different files being uploaded), the repetitions have been omitted from the table, unless there is a stark discrepancy in the prediction accuracy. Although not written explicitly under the changes, for every download and upload scenario, the file sent through the channel was different from the ones utilized in the training set. We observe the lowest results in cases where the encoding was fully changed, or the length of the tunneled subdomains has been starkly increased or decreased. We also recognize that the versatility of the open-source tools leads to lower results, due to the vast number of adjustable parameters, which can produce widely different traffic.

\begin{table*}
\centering
\caption{Classification results of the validation dataset}
\resizebox{0.74\textwidth}{!}{
\begin{tabular}{r|r|c|c|c|c|r|r}\toprule
    \multicolumn{2}{c|}{}& \multicolumn{4}{c|}{\textbf{Changes}} & \multicolumn{2}{c}{\textbf{F1-Score}}\\ \midrule
    \textbf{Malware} & \textbf{Scenario} & \multicolumn{4}{c|}{} & \textbf{Malware} & \textbf{Scenario}\\ \midrule
    Symbiote & Upload & \multicolumn{4}{c|}{} & 1.0 & 1.0 \\
    Symbiote & Download & \multicolumn{4}{c|}{} & 0.99 & 0.99 \\
    Saitama & Upload & \multicolumn{4}{c|}{} & 0.6 & 0.8 \\
    RR-PS & Upload & \multicolumn{4}{c|}{Data transfer length} & 0.17 & 0.46 \\
    RR-PS & Idle & \multicolumn{4}{c|}{} & 0.0 & 0.0 \\
    RR-PS & Download & \multicolumn{4}{c|}{} & 0.88 & 0.88 \\
    RR-PS & Download & \multicolumn{4}{c|}{Removed used RRs} & 0.88 & 1.0 \\
    RR-NET & Upload & \multicolumn{4}{c|}{} & 0.94 & 0.90 \\
    RR-NET & Upload & \multicolumn{4}{c|}{Data transfer length} & 0.74 & 0.72 \\
    RR-NET & Upload & \multicolumn{4}{c|}{Removed used domains} & 0.94 & 0.83 \\
    RR-NET & Upload & \multicolumn{4}{c|}{Encoding character set} & 0.90 & 0.84 \\
    RR-NET & Idle & \multicolumn{4}{c|}{Encoding character set} & 0.95 & 0.93 \\
    RR-NET & Download & \multicolumn{4}{c|}{} & 0.91 & 0.84 \\
    \midrule
    Iodine & Upload & \multicolumn{4}{c|}{Length to 200 chars} & 0.90 & 0.95 \\
    Iodine & Upload & \multicolumn{4}{c|}{Length to 100 chars} & 0.04 & 0.0 \\
    Iodine & Idle & \multicolumn{4}{c|}{Used encoding chars} & 0.98 & 0.98 \\
    
    Iodine & Idle & \multicolumn{4}{c|}{Utilised resource record to TXT} & 0.94 & 0.94 \\
    Iodine & Idle & \multicolumn{4}{c|}{Utilised resource record to A} & 0.99 & 0.99 \\
    Iodine & Download & \multicolumn{4}{c|}{Transfer length to 200 chars} & 0.75 & 0.75 \\
    Iodine & Download & \multicolumn{4}{c|}{Transfer length to 150 chars} & 0.54 & 0.50 \\
    DNSCat2 & Upload & \multicolumn{4}{c|}{} & 0.94 & 0.91 \\
    DNSCat2 & Upload & \multicolumn{4}{c|}{Data encoding method} & 0.03 & 0.02 \\
    DNSCat2 & Idle & \multicolumn{4}{c|}{Utilised resource record to CNAME} & 1.0 & 0.65 \\
    DNSCat2 & Idle & \multicolumn{4}{c|}{Encoding method and resource record} & 0.0 & 0.0 \\
    DNSCat2 & Download & \multicolumn{4}{c|}{} & 1.0 & 0.4 \\
    DNSCat2 & Download & \multicolumn{4}{c|}{Data encoding method} & 0.0 & 0.0 \\
    \bottomrule
\end{tabular}
}
\label{table:validation-scores}
\end{table*}

We have also validated the process on the exfiltration requests present in the \cite{Ziza2023} dataset with real traffic. These can be put into two categories based on their intentions, with the antivirus data being considered non-malicious, and the requests initialized by the authors to transfer files as malicious. Although the whitelisting of non-malicious domains that utilize DNS tunneling is possible and would prevent them from being detected as malicious, our goal is to test whether the granularity of our features can lead to them not being detected as malicious in the first place. In addition to that, the inclusion of a tunneling tool outside the ones in our training dataset allows us to examine how the classifiers handle the detection and identification of an unknown malware.

The results for both classifications are presented in Tab.~\ref{tab:results-ziza-validation}. For the identification portion of the table, only labels that were predicted more than 10 times are listed, in order to keep the visualization compact. In the case of non-malicious tunneling, we observe high accuracy in the detection classifier, with an F1-score of 0.951. A similar F1-score is shown by the identification predictions, with most of the misidentifications being attributed to the two open-source tools in our training dataset. For the McAfee tunneling channel, there were also 4 incorrect identifications as the Symbiote Dnscat2 implementation and a single misclassification as RogueRobin-Net. Additionally, the Eset exfiltration showed 1 identification as Symbiote Dnscat2.

Although the classifiers have no concept of how to categorize the DNSExfiltrator traffic, it is correctly detected as malicious in just over 75\% of the cases. However, the identification shows a much higher rate of uncertainty, with a mostly even split between the window being designated as legitimate traffic, or as being produced by Saitama, Dnscat2 or Iodine. The variance in which malware or tool the DNSExfiltrator is associated with, can be explained by the usage of different parameters (i.e., maximum domain name length and encoding) by the authors. As the true label of these requests is not within the classes of the classifier, all of these predictions considered ``incorrect'', and thus significantly reduce the overall F1-score of the identification validation.  

The Iodine traffic generated by the authors is in much smaller quantities compared to the other tools. A more in-depth analysis of the requests also showed that most of these are part of the handshake routine, separated by sporadic idle traffic. This mixture and alternation of non-exfiltration activities causes the detection classifier to have a very low success rate at predicting whether the traffic is malicious. The proportions of irregular traffic have an even starker effect on the tool identification, with none of the elements being correctly classified. Beside the identification as legitimate traffic, there are also 3 classifications as RogueRobin-Net and 1 as RogueRobin-PS.

\begin{table}
\centering
\resizebox{1\textwidth}{!}{
    \begin{tabular}{r| c c c |c c c c c}
        & & \multicolumn{2}{c|}{Predicted Detection} & \multicolumn{4}{c}{Predicted Identification}\\ \midrule
        Tool & Actual & {Malic.} & Non-malic. & {Saitama} & {Dnscat2} & {Iodine} & {Legitim.} & {Others}\\ \midrule
        Eset & Non-malic. & 149 & 11,627 & 0 & 52 & 73 & 11,650 & 1\\
        McAfee & Non-malic. & 48 & 880 & 0 & 2 & 14 & 907 & 5\\
        Iodine & Malic. & 4 & 25 & 0 & 0 & 0 & 25 & 4\\
        DNSExfilt. & Malic. & 3,661 & 1,065 & 997 & 1,101 & 1,554 & 1,074 & 0\\
        \bottomrule
    \end{tabular}
  {\caption{\label{tab:results-ziza-validation}Classification results for the validation of third-party traffic.}}
}
\end{table}

While there is no full overlap of the investigated tools, we have also evaluated our approach on the dataset provided by the authors of the GraphTunnel \cite{GraphTunnel} framework. The dataset contains multiple recordings of both \textit{Iodine} and \textit{dnscat2} traffic, with the \textit{Iodine} file names suggesting alternation between the used resource records. However, no specifications are given for any of the other parameters. The captured \textit{dnscat2} traffic was detected and identified correctly by our classifiers with an accuracy of $0.9$. Although the performed actions are not given, our behavior classifier overwhelmingly suggests the presence of a file upload, together with idle traffic. The \textit{Iodine} traffic identification was less successful, as the classifier recognized the tool about half the time in the files designated as \textit{NULL} and \textit{TXT} resource records, as well as the recording named \textit{private}. The rest of the recordings, indicated as \textit{A}, \textit{MX}, \textit{CNAME} and \textit{SRV} had a prediction rate of less than $0.05$. According to our behavior classifier, the tool provides download functionality. Manual inspection of the recordings shows the main difference between them appears to be in what actions were performed, with the \textit{NULL}, \textit{TXT} and \textit{private} recordings exhibiting traffic very similar to our \textit{Iodine-Download} scenario, and the remainder being of unclear behavior. The provided normal traffic was correctly detected as non-malicious with a rate of $0.96$, ascertaining a low false-positive rate on previously unseen traffic.

\subsection{Result Comparison}
Although in Sect.~\ref{sect:relwrk} we have listed multiple other tools that cover tunneling detection as their main goal, none of them have made their models and feature extraction partially or fully available. For this reason, we cannot fairly evaluate their approaches on the dataset containing real malware samples provided by us, and therefore have chosen to compare the results to the ones that have been provided in each work. Tab.~\ref{tab:related-work-comparison} compares the results of our approach with the related work.

\begin{table}
\resizebox{1\textwidth}{!}{
    \begin{tabular}{r|c|c|c|c|c}
    \toprule
        \multirow{2}{*}{\textbf{Tool}} & \textbf{Malware} & \textbf{Open-Source} & \textbf{Tunnel} & \textbf{Tunnel} & \textbf{Behavior} \\
        & \textbf{Count} & \textbf{Count} & \textbf{Detection} & \textbf{Identification} & \textbf{Identification} \\
        \midrule
        \textbf{Domainator} (our approach) & \MalwareCount{} & \ToolsCount{} & 0.966 & 0.964 & 0.857 \\
        \textbf{GraphTunnel} (2024) \cite{GraphTunnel} & 0 & 7 & 1.000 & 0.986 & \cdpNA \\
        \textbf{Žiža \emph{et al.}} (2023) \cite{Ziza2023} & 0 & 2 & 0.998 & \cdpNA & \cdpNA \\
        \textbf{Alkasassbeh \emph{et al.}} (2023) \cite{alkasassbeh} & 0 & 1 & 0.97 & \cdpNA & \cdpNA \\
        \textbf{Buczak \emph{et al.}} (2016) \cite{buczak} & 0 & 4 & 0.95 & \cdpNA & \cdpNA \\
        \bottomrule
    \end{tabular}
  {\caption{Result comparison with other similar tools}\label{tab:related-work-comparison}}
  }
\end{table}

While our detection approach achieves a slightly lower detection score, none of the other tools have attempted to detect \emph{real} malware, i.e., they solely covered open source DNS tunneling tools. Further, our approach is the first to perform an identification of real-world malware tool that produced the detected traffic (GraphTunnel also identified tools but did not cover real-world malware). Given this more challenging task, our detection score of 0.964 can be considered promising. Furthermore, \ToolName~is the first attempt to identify specific \emph{actions} taken by the malware based on its DNS tunneling activities, and thus no comparison can be drawn to other tools. The achieved score of 0.857 is acceptable for this specific task.

\section{Discussion}\label{sect:disc}

\ToolName{} differentiates between certain malware samples and their actions, based on DNS tunneling behavior.

Our method solely relies on the metadata analysis of subdomain strings, i.e., it works independently of other DNS characteristics that are utilized by malware, such as the transferring of commands and payloads in \texttt{TXT} answer records, which provides a high embedding capacity.

\ToolName{} is tailored to aid the classification and attribution of upcoming malware families. Therefore, DNS traffic of new malware must be recorded, e.g., in our testbed. If the traffic matches certain sequential patterns identified by our research, it might be linked to the specific type of malware that we associate with the particular pattern. As discussed in Sect.~\ref{sub:scenario-ident}, we are unable to successfully identify some idle scenarios, which is caused by the idle behavior of the malware continuously sending identical requests and thus closely resembling legitimate traffic. During our evaluations, we observed only PowerShell RogueRobin and Saitama being incorrectly classified for those reasons, but future malware with such idle behavior could result in similar misclassifications. Although the per window false-positive rate of the non-malicious traffic is already low, adding a per domain threshold based on malicious window count can further lower the per domain false-positive-rate.

\textit{Limitations} We investigated \TotalCount{} malware samples and tools that exploit subdomain strings in the DNS protocol. Other or future malware samples could apply covert channel techniques for DNS in alternative ways, e.g., based on different hiding patterns~\cite{csur}. Such malware would most likely require us to adjust the set of selected DNS features for appropriate detection. Moreover, future versions of the analyzed malware might change their behavior or encoding strategy for subdomains. This could influence detectability and would also require adjustments of our features.
The identification of future malware outside of our current set would not be possible without retraining the classifier with the traffic from the new malware, as the classifier would have no knowledge of it. Versatile tools like \textit{Iodine} would also benefit from multiple parameter setups being added to the dataset, due to the vastly different tunneling traffic it can produce based on those settings, as seen in our validation results. 
Finally, we experimented solely with a limited set of simple classifiers  and evaluating alternative, more complex machine learning methods might improve our results.

\textit{Provision of Data} Our traffic dataset is available for review (it will be released under a BSD license): \url{https://anonymous.4open.science/r/dataset-7ED5}

\section{Conclusion}\label{sect:concl}

We introduced a sequence-based detection and differentiation approach for malware using DNS-based covert channel techniques. 
We utilize only the subdomain portions of the domains sent in the DNS requests as a vector to calculate statistical metrics. These features were used to train a generic Random Forest classifier, which can be used for both the \emph{detection of DNS tunneling traffic} and the \emph{identification of the malware}.
Our approach is even able to \emph{identify the action undertaken by the malware} (e.g., file upload or download), utilizing the same statistical features.
In future work, we plan to extend our research to additional protocols of the TCP/IP protocol suite, especially to HTTP(S).

\begin{credits}
\subsubsection{\ackname} Parts of the research work have been funded by the ATTRIBUT project, \emph{Agentur für Innovation in der Cybersicherheit}, Germany, program ``Existenzbe\-drohenden Risiken aus dem Cyber- und Informationsraum – Hochsicherheit in sicherheitskritischen und verteidigungsrelevanten Szenarien'' (HSK), \emph{https://www.cyberagen- tur.de/tag/hsk/}. Project website: \emph{https://attribut.cs.uni-magdeburg.de}.

\subsubsection{\discintname}
The authors have no competing interests to declare that are relevant to the content of this article.
\end{credits}

\bibliographystyle{plain}
\bibliography{quellen}

\end{document}